\documentclass[sigconf,natbib=true,screen=true]{acmart}
\usepackage{amsmath}
\usepackage{booktabs} % For formal tables
\usepackage{adjustbox}
\usepackage{algorithm, algpseudocode}
\usepackage{amsmath}
\usepackage{graphics}
\usepackage{epsfig}
\usepackage{graphicx}
\usepackage{xcolor}
\usepackage[skip=5pt]{caption}
\usepackage{subfigure}
\usepackage{balance}
\usepackage{multirow}
\usepackage{mathrsfs}
\usepackage{acronym}
\usepackage{placeins}
\usepackage{xcolor}
\usepackage[inline]{enumitem}

\usepackage{url}
\usepackage{xspace}
\usepackage{bbm}

\newcommand\BibTeX{B{\sc ib}\TeX}

\AtBeginDocument{%
  \providecommand\BibTeX{{%
    \normalfont B\kern-0.5em{\scshape i\kern-0.25em b}\kern-0.8em\TeX}}}

\acrodef{CF}{collaborative filtering}
\acrodef{MIAs}{membership inference attacks}
\acrodef{MIA}{membership inference attack}
\acrodef{MF}{matrix factorization}
\acrodef{LFM}{latent factor model}
\acrodef{NCF}{neural collaborative filtering}
\acrodef{DL-MIA}{Debiasing Learning for Membership Inference Attacks against recommender systems}
\acrodef{VAE}{variational auto-encoder}
\acrodef{ItemBase}{Item-based collaborative filtering}
\acrodef{MLP}{multilayer perceptron}
\acrodef{vMF}{von Mises Fisher}

\newcommand{\header}[1]{\vspace*{1mm}\noindent\textbf{#1.}}
\newcommand{\changed}[1]{\textcolor{blue}{#1}}

\definecolor{french_blue}{RGB}{0, 112, 182}

\author{Zhaochun Ren*}
\orcid{0000-0002-9076-6565}
\affiliation{%
  \institution{Shandong University}
  \city{Qingdao}
  \country{China}
}
\email{zhaochun.ren@sdu.edu.cn}

\author{Na Huang*}
\orcid{0009-0003-3844-3237}
\affiliation{%
  \institution{Shandong University}
  \city{Qingdao}
  \country{China}
}
\email{hn.z@mail.sdu.edu.cn}

\author{Yidan Wang}
\orcid{0009-0002-5772-5910}
\affiliation{%
  \institution{Shandong University}
  \city{Qingdao}
  \country{China}
}
\email{yidanwang@mail.sdu.edu.cn}

\author{Pengjie Ren}
\orcid{0000-0003-2964-6422}
\affiliation{%
  \institution{Shandong University}
  \city{Qingdao}
  \country{China}
}
\email{jay.ren@outlook.com}

\author{Jun Ma}
\orcid{0000-0003-0203-4610}
\affiliation{%
  \institution{Shandong University}
  \city{Qingdao}
  \country{China}
}
\email{chenzhumin@sdu.edu.cn}

\author{Jiahuan Lei}
\orcid{0000-0002-5170-8645}
\affiliation{%
  \institution{Meituan}
  \city{Beijing}
  \country{China}
}
\email{leijiahuan@meituan.com}

\author{Xinlei Shi}
\orcid{0000-0002-0733-5757}
\affiliation{%
  \institution{Meituan}
  \city{Beijing}
  \country{China}
}
\email{shixinlei@meituan.com}

\author{Hengliang Luo}
\orcid{0000-0001-8597-8873}
\affiliation{%
  \institution{Meituan}
  \city{Beijing}
  \country{China}
}
\email{luohengliang@meituan.com}

\author{Joemon M Jose}
\orcid{0000-0001-9228-1759}
\affiliation{%
  \institution{University of Glasgow}
  \city{London}
  \country{UK}
}
\email{Joemon.Jose@glasgow.ac.uk}

\author{Xin Xin$^{\dagger}$}
\orcid{0000-0001-6116-9115}
\affiliation{%
  \institution{Shandong University}
  \city{Qingdao}
  \country{China}
}
\email{xinxin@sdu.edu.cn}

\makeatletter
\def\authornotetext#1{
\if@ACM@anonymous\else
    \g@addto@macro\@authornotes{
    \stepcounter{footnote}\footnotetext{#1}}
\fi}
\makeatother
\authornotetext{Equal contribution.}
\authornotetext{Corresponding author.}

\def\authors{Zhaochun Ren, Na Huang, Yidan Wang, Pengjie Ren, Jun Ma, Jiahuan Lei, Xinlei Shi, Hengliang Luo, Joemon M.Jose, and Xin Xin}
\renewcommand{\shortauthors}{Ren et al.}

\copyrightyear{2023}
\acmYear{2023}
\setcopyright{acmlicensed}\acmConference[SIGIR '23]{Proceedings of the 46th
International ACM SIGIR Conference on Research and Development in
Information Retrieval}{July 23--27, 2023}{Taipei, Taiwan}
\acmBooktitle{Proceedings of the 46th International ACM SIGIR Conference on
Research and Development in Information Retrieval (SIGIR '23), July 23--27,
2023, Taipei, Taiwan}
\acmPrice{15.00}
\acmDOI{10.1145/3539618.3591656}
\acmISBN{978-1-4503-9408-6/23/07}

\renewcommand{\shortauthors}{Zhaochun Ren et al.}

\begin{document}
\begin{sloppypar}

% \fancyhead{}
% \fancyfoot{}

\title{Contrastive State Augmentations for Reinforcement Learning-Based Recommender Systems}

% !TEX root = ../main.tex

\begin{abstract}
Learning reinforcement learning (RL)-based recommenders from historical user-item interaction sequences is vital to generate high-reward recommendations and improve long-term cumulative benefits. However, 
existing RL recommendation methods encounter difficulties
\begin{enumerate*}[label=(\roman*)]
\item to estimate the value functions for states which 
are not contained in the offline training data, and 
\item to learn effective state representations from user implicit feedback due to the lack of contrastive signals.
\end{enumerate*}

In this work, we propose \textbf{c}ontrastive \textbf{s}tate \textbf{a}ugmentations (CSA) for the training of RL-based recommender systems. 
To tackle the first issue, we propose four state augmentation strategies to enlarge the state space of the offline data. The proposed method improves the generalization capability of the recommender by making the RL agent visit the local state regions and ensuring the learned value functions are similar between the original and augmented states. For the second issue, we propose introducing contrastive signals between augmented states and the state randomly sampled from other sessions to improve the state representation learning further. 

To verify the effectiveness of the proposed CSA, we conduct extensive experiments on two publicly accessible datasets and one dataset collected from a real-life  e-commerce platform. We also conduct experiments on a simulated environment as the online evaluation setting. 
Experimental results demonstrate that CSA can effectively improve recommendation performance. 
\end{abstract}



\maketitle

%\acresetall

% !TEX root = ../main.tex

\section{Introduction}

\label{sec:Introduction}
Sequential recommendation shows promising improvement in predicting users' dynamic interests. It has been successfully deployed to provide personalized services in various application scenarios, such as e-commerce platforms, social networks, and lifestyle apps ~\citep{xiao2007commerce,jamali2010matrix,wang2018billion,schafer2001commerce}. 
Recent advances in deep neural networks inspire the recommendation community to adopt various kinds of models for modelling user-item interaction sequences, e.g., Markov chains~\citep{rendle2010factorization,rendle2010factorizing}, recurrent neural networks~\citep{hidasi2015session,hidasi2016parallel}, convolutional neural networks~\citep{tang2018personalized,yan2019cosrec}, and attention-based methods~\citep{kang2018self,sun2019bert4rec}. 
These methods are used to characterize the correlation among item transitions and learn representations of user preference. 
Although these methods have shown promising performance, they are usually trained with pre-defined supervision signals, such as next-item or random masked item predictions. Such supervised training of recommenders can result in sub-optimal performance since the model is purely learned by a loss function based on the discrepancy between model prediction and the supervision signal. The supervised loss may not match the expectation from the perspective of service providers, e.g., improving long-term benefits or promoting high-reward recommendations.

To maximize cumulative gains with more flexible reward settings, reinforcement learning (RL)
has been applied to the sequential recommendation task.
An obstacle in applying existing RL methods to recommendation is that conventional RL algorithms belong to a fundamentally online learning paradigm. Such a learning process of online RL involves iteratively collecting experiences by interacting with the user. However, this iterative online approach would be costly and risky for real-world recommender systems.
An appealing alternative is utilizing offline RL methods, which target on learning policies from logged data without online explorations affecting the user experience ~\citep{fujimoto2019off,kumar2020conservative}. Although there exists some research focusing on offline RL ~\citep{gao2022value,xiao2021general,zhang2022text}, how to design appropriate offline RL solutions for the sequential recommendation task remains an open research challenge due to the following limitations:    

\begin{itemize}[leftmargin=*,nosep]
\item Potentially huge user state space limits the generalization capability of the offline RL algorithms. Since offline RL algorithms aim to learn policies without online explorations, these algorithms can only investigate the state-action pairs that occurred in the logged training data. During the model inference, there could be new and out-of-distribution user states. Besides, the state transition probability could also be different from the offline data. Consequently, the RL recommendation agent could suffer from severe distribution shift problems, resulting in inaccurate estimating of the value functions for the data that is not in the same distribution as the offline training set.

\item The lack of contrastive signals makes the RL agent fail to learn effective state representations. Modern recommender systems are usually trained based on implicit feedback data, which only contains positive feedback (e.g., clicks and purchases). The lack of negative feedback could lead to a situation in which the RL agent cannot know which state is bad or which action should be avoided for a given state. Given the sparse user-item implicit interactions, how to improve the data efficiency to learn effective state representations still needs to be investigated to improve the performance of RL-based recommender systems further.  
\end{itemize}

To address the above issues, we propose a simple yet effective training framework, which explores contrastive state augmentations (CSA) for RL-based recommender systems. 
More precisely, given an input item sequence, we use a sequential recommendation model to map the sequence into a hidden state, upon which we stack two final output layers. One is trained with the conventional supervised cross-entropy loss, while the second is trained through double Q-learning ~\citep{hasselt2010double}.
To tackle the first limitation, we propose four state augmentation strategies for the RL output layer to enlarge the state space of the offline training data. Such an approach smooths the state space by making the RL agent explicitly visit the local state regions and ensuring the learned value functions are similar among the original sequence state and the augmented states since small perturbations in the given observation should not lead to drastically different value functions.
For the second limitation, we use contrastive learning on the RL output layer to "pull" the representations of different augmentations of the same sequence state towards each other while "pushing" the state representations from different sequences apart. 

Finally, we co-train the supervised loss, the RL loss over the original and augmented states, and the contrastive loss from the logged implicit feedback data.
To verify the effectiveness of our method, we implement CSA on three state-of-the-art sequential recommendation models and conduct experiments on two publicly accessible benchmark datasets and one dataset collected from a real-life  e-commerce platform. We also conduct experiments on a simulated environment to verify whether CSA can help to improve cumulative gains on multi-round recommendation scenarios. Experimental results demonstrate the superior performance of the proposed CSA. 

To summarize, our main contributions are as follows:
\begin{itemize}[leftmargin=*,nosep]
\item  We propose a simple yet effective contrastive state augmentation method to train RL-based recommender systems. The proposed approach can be seen as a general framework, which can be incorporated with several off-the-shelf sequential models.

\item We propose four state augmentation strategies to improve the generalization capability of the RL recommendation agent. Besides, we propose a contrastive loss to improve the state representation learning.

\item Extensive experiments conducted on three real-world datasets and one simulated online environment demonstrate the effectiveness of our proposed approach.
\end{itemize}

% !TEX root = ../main.tex

\section{Related work}
\label{sec:Related works}
This section provides a literature review regarding sequential recommendation, reinforcement learning, and contrastive learning.

\subsection{Sequential Recommendation}
Early sequential recommendation methods mainly rely on the Markov Chain (MC). MC-based methods estimate an item-item transition probability matrix and utilize it to predict the next item given a user's last interactions. \citet{rendle2010factorizing} combined matrix factorization and the first-order MC to capture both general and short-term user interests.  
Methods with high-order MCs that consider longer interaction sequences were also developed in ~\citep{he2016vista,he2016fusing}. 
% A shortcoming of MC-based models is that it is challenging to learn long-term dependencies because MC-based models assume that the next state is only related to the previously nearest state. Although high-order MC can address this challenge, it consumes high computational costs. 
Many deep learning-based approaches have recently been proposed to model the user-item interaction sequences more effectively.
%As a result, plenty of deep learning-based approaches have been proposed to model the interaction sequences more effectively.
\citet{hidasi2015session} firstly introduced Gated Recurrent Units (GRU) into the session-based recommendation task,  and a surge of following variants modified this model by incorporating pair-wise loss function ~\citep{hidasi2016parallel}, copy mechanism ~\citep{ren2019repeatnet}, memory networks ~\citep{huang2019taxonomy,huang2018improving} and hierarchical structures ~\citep{quadrana2017personalizing}, etc. 
However, RNN-based methods assume that the adjacent items in a session have sequential dependencies, which may fail to capture the skip signals. \citet{tang2018personalized} and \citet{yuan2019simple} proposed to utilize convolutional neural networks (CNN) to model sequential patterns from local features of the embeddings of previous items. \citet{kang2018self} proposed exploiting the well-known Transformer ~\citep{vaswani2017attention} in the sequential recommendation field.
%to capture users' long-term preferences effectively. 

% BERT4Rec~\citep{sun2019bert4rec} proposed to model user behaviour sequences with a bidirectional self-attention network through the Cloze task. 

Regarding the learning task, most existing sequential recommendation methods utilize the next-item prediction task \cite{hidasi2015session,yuan2019simple,kang2018self,Cheng2023MBGCN}. Besides, self-supervised learning has demonstrated its effectiveness in representation learning by constructing training signals from raw data other than external labels.
\citet{sun2019bert4rec} proposed to use the task of predicting random masked items to train sequential recommenders.
\citet{zhou2020s3} proposed four auxiliary self-supervised tasks to maximize the mutual information among attributes, items, and sequences. 
\citet{xia2021self} proposed a self-supervision task to maximize the mutual information between sequence
representations learned from hypergraphs. 

Despite the advances of the above methods, they are trained to minimize the discrepancy between model predictions and pre-defined (self-)supervision signals. Such learning signals may not match the recommendation expectation, e.g., improving cumulative gains in one interaction session.

% However, all these approaches are purely learned by a loss function defined on the discrepancy between model predictions and the self-supervision signal. Such a loss may not match the expectations from the perspective of recommendation systems.

\begin{figure*}[t]
    \centering
    \includegraphics[width=0.85\linewidth]{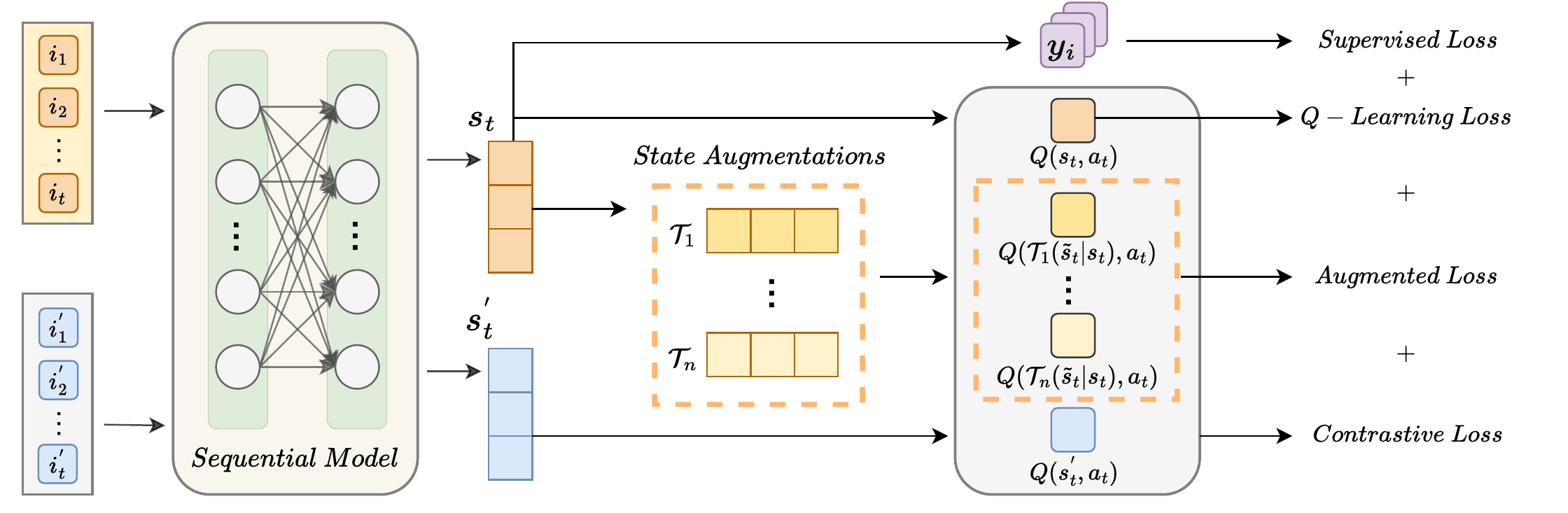}
    \caption{The overview of CSA. $\{i_1,i_2,...,i_t\}$ denotes the original input sequence. $\{\mathcal{T}_1,...,\mathcal{T}_n\}$ denotes state augmentations. The base sequential model is trained jointly through the supervised loss, the RL loss over the original state (i.e, Q-learning loss) and the augmented states (i.e., augmented loss), and the contrastive loss over the comparison between $\{i'_1,i'_2,...,i'_t\}$ which is a sampled sequence from another session.}
    \label{fig:the framework.}
    \vspace{-0.3cm}
\end{figure*}

\subsection{Reinforcement Learning}
RL has shown promising improvement to increase cumulative gains in the long-run. Conventional RL methods can be categorized into on-policy methods (e.g., policy gradient ~\citep{sutton1999policy}) and off-policy methods (e.g., Q-learning ~\citep{watkins1989learning,mnih2015human}, actor-critic ~\citep{konda1999actor,haarnoja2018soft}). On-policy methods target on learning policies through real-time interactions with the environment. Off-policy methods exploit a replay buffer to store past experiences and thus improve the data efficiency of the RL algorithm. Both on-policy and off-policy methods need to perform online explorations to collect the training data\footnote{We follow the definition of ~\citep{levine2020offline}   that both on-policy methods (e.g., policy gradient) and off-policy methods (e.g., Q-learning and actor-critic) are online RL methods.}. On the contrary, offline RL ~\citep{levine2020offline} aims to train the agent from fixed data without new explorations. Recent research has started to investigate the offline RL problem ~\citep{kumar2019stabilizing,fujimoto2019off,gao2022value}, e.g., addressing the over-estimation of value functions ~\citep{kumar2019stabilizing,fujimoto2019off} or learning from expert demonstrations ~\citep{gao2022value}.

RL has recently been introduced into recommender systems because it considers users' long-term engagement~\citep{zhao2018recommendations, zou2020pseudo}. 
~\citet{zhao2018deep} proposed to generate list-wise and page-wise recommendations using RL.
% The MDP-Based CF model approximates a partial observable MDP (POMDP) by using a finite rather than unbounded history window to define the current state~\citep{shani2005mdp}. To reduce the high computational and representational complexity of POMDP, three strategies have been developed: value function approximation~\citep{hauskrecht1997incremental}, policy-based optimization~\citep{poupart2004vdcbpi}, and stochastic sampling~\citep{kearns2002sparse}.
To address the problem of distribution shift under the off-policy settings, ~\citet{chen2019top} proposed utilizing propensity scores to perform the off-policy correction. However, such methods suffer from high variances in the estimated propensity score.
Model-based RL approaches ~\citep{chen2019generative,zou2019reinforcement} firstly build a model to simulate the environment. The policy is then trained through interactions with the constructed simulator. ~\citet{xin2020self} proposed to infuse self-supervision signal to improve the training of the RL agent. Besides, contextual information has been considered to enhance the RL process for recommendations. For example, ~\citet{xian2019reinforcement} proposed a policy gradient
approach to extract paths from knowledge graphs and regard these paths as the interpretation of the recommendation process. ~\citet{wang2020kerl} fused historical and future knowledge to guide RL-based sequential recommendation.

Given the large user state space, how to perform effective offline learning for RL-based sequential recommendation to maximize long-term cumulative benefits remains an open research challenge. 

\subsection{Contrastive Learning}
Recently, contrastive learning achieves remarkable successes in various fields, such as speech processing~\citep{oord2018representation}, computer vision~\citep{chen2020simple}, as well as natural language processing~\citep{gao2021simcse,wu2020clear}. By maximizing mutual information among the positive transformations of the data itself while improving discrimination ability to the negatives, it discovers the semantic information shared by different views and gets high-quality representations. 
As a result, various recommendation methods ~\citep{zhou2020s3,wu2021self,xie2022contrastive} employ contrastive learning to optimize the representation learning. Specifically, ~\citet{zhou2020s3} proposed to adopt a pre-train and fine-tuning strategy and utilize contrastive learning during pre-training to incorporate correlations among item meta information. ~\citet{wu2021self} proposed a multi-task framework with contrastive learning to improve the graph-based collaborative filtering methods.
~\citet{xie2022contrastive} proposed to utilize a contrastive objective to enhance user representations. They used item crop, item mask, and item reorder as data augmentation approaches to construct contrastive signals. \citet{qiu2022contrastive} performed contrastive self-supervised learning based on dropout.

Different from existing works which focus on utilizing contrastive learning under the (self-)supervised paradigm, our research explores contrastive signals to improve the representation learning for RL-based recommenders.

% Our approach explores contrastive signals to improve the representation learning of RL-based recommenders. Furthermore, this contrasts existing works, focusing on utilizing contrastive learning under the (self-)supervised paradigm.
% !TEX root = ../main.tex

\section{Method}
\label{sec:Method}
In this section, we first formulate the recommendation task of our research (Sec.~\ref{subsec:Problem formulation}).
Then we introduce the primary settings of the RL setup (Sec.~\ref{subsec:Reinforcement learning}).
The state augmentation is detailed in section \ref{subsec:Data Augmentations}.
Finally, the contrastive learning for RL-based recommender is described in section \ref{subsec:Self-supervision in Offline RL}. Figure~\ref{fig:the framework.} shows the overview of CSA.

\subsection{Task Formulation}
\label{subsec:Problem formulation}
Assume that we have a set of items, denoted by $\mathcal{I}$ where $i \in \mathcal{I}$ denotes an item. Then a user-item interaction sequence can be represented as $x_{1:t}=\{i_1,...,i_j,...,i_{t}\}$, where $t$ is the length of interactions and $i_j$ is the item that the user interacted with at the $j$-th time step. 
The goal of conventional sequential recommendation is to predict the next item that the user is likely to interact with (i.e., $i_{t+1}$) given the sequence of previous interactions $\mathit{x}_{\mathit{1:t}}$.

Generally speaking, one can use a sequential model $G(\cdot)$ (e.g., GRU) to map the input sequence $x_{1:t}$ into a hidden representation $\mathbf{s}_{t} = \mathit{G}\left(x_{\mathit{1:t}}\right)$, which can be seen as a general encoding process. Based on the hidden representation, one can utilize a decoder to map the representation into a ranking score for the next time step as $\mathbf{y}_{t+1} = f\left(\mathbf{s}_{\mathit{t}}\right) =[y_1,y_2,...,y_{|\mathcal{I}|}] \in \mathbb{R}^{|\mathcal{I}|}$. $f$ is usually defined as a simple, fully connected layer or the inner product with candidate item embeddings. During the model inference, top-$k$ items with the highest ranking scores are selected as the recommended items. Note that in this research, we expect the recommended items can lead to high-reward user feedback (e.g., purchases in the e-commerce scenario) or better cumulative benefits in the long-run.

\subsection{Reinforcement Learning Setup}
\label{subsec:Reinforcement learning}
From the perspective of RL, a sequential recommendation can be framed as a Markov Decision Process (MDP), in which the recommendation agent interacts with the environments (users) by sequentially recommending items to maximize the cumulative rewards. The MDP can be represented by a tuple of $\left(\mathcal{S}, \mathcal{A},\rho,\mathit{r}, \gamma \right)$: 
\begin{itemize}[leftmargin=*,nosep]
    \item $\mathcal{S}$ is the state space of the environment. It is modeled by the user's historical interactions with the items. Specifically, the state of the user can be denoted as the hidden representation from the sequential model: $\mathbf{s}_{t}=G\left(\mathit{x}_{\mathit{1:t}}\right) \in \mathcal{S}$.
    \item $\mathcal{A}$ is the action space. At each time step $t$, the agent selects an item from the candidate pool and recommends it to the user. In offline training data, we get the action at timestamp $t$ from the logged user-item interaction (i.e. $a_t=\mathit{i}_{{t+1}} \in \mathcal{I}$)\footnote{In this paper, we follow the top-$k$ recommendation paradigm. We leave the recommendation of a set of items as one of our future work.}.
    \item $\rho\left(\mathbf{s}_{t+1}|\mathbf{s}_{t}, a_{t}\right)$ is the transition function given the current state and the taken action, describing the distribution of the next state.
    \item $r\left(\mathbf{s}_{t},a_t\right)$ is the immediate reward if action $a_t$ is taken under $\mathbf{s}_t$.
    \item $\gamma$ is the discount factor for the future reward with $\gamma \in [0, 1]$.
\end{itemize}
The RL algorithm seeks a target policy $\pi_{\theta}\left(a|\mathbf{s}\right)$, where $\theta$ denotes the parameters, to translate the user state $\mathbf{s} \in \mathcal{S}$ into a distribution over actions $a \in \mathcal{A}$.
The policy is trained to maximize the expected cumulative discount reward in the MDP as
\begin{equation}
    \mathop\text{max}_{\substack{\pi_{\theta}}} \mathop\mathbb{E}_{\tau\sim\pi_{\theta}}\left[\mathcal{R}\left(\tau\right)\right], \text{where }  \mathcal{R}\left(\tau\right)=\sum_{t=0}^{|\tau|} \gamma^{\mathit{t}} \mathit{r}\left(\mathbf{s}_{\mathit{t}}, \mathit{a}_{\mathit{t}} \right).
\end{equation}
$\tau$ denotes a trajectory of $(\mathbf{s}_t,a_t,\mathbf{s}_{t+1})$ which is sampled according to the target policy $\pi_{\theta}$.  
To train the target policy, we adopt the value-based Q-learning algorithm \citep{silver2016mastering} in this work. We don't utilize the policy gradient method since it is based on the on-policy update from the Monte-Carlo sampling and needs plenty of online explorations, which may affect the user experience. Although~\citet{chen2019top} proposed to utilize inverse propensity scores to correct the offline data distribution, such kinds of methods could suffer from severe high variance. 

Q-learning utilizes policy evaluation to calculate the Q-value, which is defined as the estimation of the cumulative reward for an action $a_t$ given the state $\mathbf{s}_{t}$. 

The Q-value is trained by minimizing the one-step time difference (TD) error:
\begin{equation}
    \mathcal{Q}_{\mathit{i+1}}\leftarrow \mathop\text{arg min}_{\substack{\mathcal{Q}}}\mathbb{E}_{\pi_{i}}\left[\mathit{r}_{\mathit{t}}+\gamma\mathcal{Q}_{\mathit{i}}(\mathbf{s}_{\mathit{t+1}},\mathit{a}_{\mathit{t+1}})-\mathcal{Q}_{\mathit{i}}(\mathbf{s}_{\mathit{t}},\mathit{a}_{\mathit{t}})\right],
\end{equation}
where $\mathcal{Q}_{\mathit{i}}$ is the Q-value in the $i$-th step of policy evaluation. $\pi_{\mathit{i}}$ is the target policy in the $i$-th update step.
When learning from logged offline data, the policy evaluation can be reformulated as
\begin{equation}
    \mathcal{Q}_{\mathit{i+1}}\leftarrow \mathop\text{arg min}_{\substack{\mathcal{Q}}}\mathbb{E}_{(\mathbf{s}_{\mathit{t}},a_t)\sim\mathcal{D}}\left[\mathit{r}_{\mathit{t}}+\gamma\mathcal{Q}_{\mathit{i}}(\mathbf{s}_{\mathit{t+1}},\mathit{a}_{\mathit{t+1}})-\mathcal{Q}_{\mathit{i}}(\mathbf{s}_{\mathit{t}},\mathit{a}_{\mathit{t}})\right],
\end{equation}
where $\mathcal{D}$ is the collected historical data. % collected by behavior policy $\beta$.
The policy is then improved to maximize the state-action Q-values of performing an action $\mathit{a}_{\mathit{t}}$ given $\mathbf{s}_{\mathit{t}}$, also known as policy improvement:
\begin{equation}
    \pi_{\mathit{i+1}}\leftarrow \mathop\text{arg max}_{\substack{\pi}}\mathbb{E}\left[\mathcal{Q}(\mathbf{s}_{\mathit{t}},\pi_{\mathit{i}}(\mathit{a}_{\mathit{t}}|\mathbf{s}_{\mathit{t}}))\right].
\end{equation}

In deep Q-learning, the Q-values are parameterized using neural networks (with parameter $\theta$) and trained using gradient descent.

In this work we utilize double Q-learning to further enhance the learning stability and follow \cite{xin2020self} to combine supervised learning and RL through a shared base sequential model. More specifically, given the base model $G(\cdot)$ and the input item sequence $\mathit{x}_{\mathit{1:t}}$, the hidden representation can be formulated as $\mathbf{s}_{\mathit{t}}=\mathit{G}(\mathit{x}_{\mathit{1:t}})$, which is then fed into a fully connected layer. The layer outputs a $|\mathcal{I}|$-dimensional vector $\mathbf{y}_t$, denoting the prediction ranking scores over the item set:
\begin{equation}
\mathit{\textbf{y}_t} =\phi(\mathit{\textbf{W}_{\mathit{s}}}\mathbf{s}_{\mathit{t}}+\mathit{\textbf{b}}_{\mathit{s}}),
\end{equation}
where $\phi$ denotes the activation function. $\mathit{\textbf{W}}_{\mathit{s}}$ and $\mathit{\textbf{b}}_{\mathit{s}}$ are trainable parameters. Then the supervised training loss can be defined as the cross-entropy over the ranking distribution:
\begin{equation}
\label{eq:cne loss}
    \mathcal{L}_{\mathit{s}} = -\sum_{i=1}^{\mathcal{|I|}}\mathit{Y}_{\mathit{i}}\log(\mathit{p}_{\mathit{i}}), \text{where }   \mathit{p}_{\mathit{i}} = \frac{\mathit{e}^{y_{\mathit{i}}}}{\sum_{\mathit{i}^{\mathit{'}}=1}^{|\mathcal{I}|}\mathit{e}^{y_{\mathit{i}^{\mathit{'}}}}},
\end{equation}
where $\mathit{Y}_{\mathit{i}}$ indicates the ground truth and $\mathit{Y}_{\mathit{i}}=1$ denotes the user interacted with the $i$-th item, otherwise $\mathit{Y}_{\mathit{i}}=0$.

Regarding the Q-learning network, we stack the other fully connected layer upon the hidden representation $\mathbf{s}_t$.
Then we calculate the Q-values as
\begin{equation}
    \mathcal{Q}(\mathbf{s}_{\mathit{t}},\mathit{a}_{\mathit{t}}) = \phi(\mathit{\textbf{W}_{\mathit{r}}}\mathit{\textbf{s}}_{\mathit{t}}+\mathit{\textbf{b}}_{\mathit{r}}),
\end{equation}
where $\mathit{\textbf{W}}_{\mathit{r}}$ and $\mathit{\textbf{b}}_{\mathit{r}}$ are trainable parameters of the RL layer. The Q-learning loss can be defined as
\begin{equation}
\label{eq:q-loss}
    \mathcal{L}_{q}=({\mathit{r}(\mathbf{s}_{\mathit{t}},\mathit{a}_{\mathit{t}})+\gamma\max_{\mathit{a}'}\mathcal{Q}(\mathbf{s}_{\mathit{t+1}},\mathit{a}')-\mathcal{Q}(\mathbf{s}_{\mathit{t}},\mathit{a}_{\mathit{t}})})^2.
\end{equation}

\subsection{State Augmentations}
\label{subsec:Data Augmentations}
Due to the fact that the recommender is trained on the historical data without online explorations, the agent can only investigate the state-action pairs that occurred in the offline training data. As a result, the Q-network may fail to estimate the value functions for unseen states. To this end, we propose to use state augmentations to learn better value functions during the training stage. 
Based on the described RL setup, we then discuss the design of the transformations for a given state, which combines local perturbations to the states with similar value function estimation to enhance the generalization capability  of the RL agent. 
Augmentations such as rotations, translations, colour jitters, etc., are commonly used to increase the variety of data points in the field of computer vision. Such transformations can preserve the semantics of the image after the transformation. However, in the recommendation scenario, a state augmentation strategy that is too aggressive may end up hurting the RL agent since the reward for the original state may not coincide with the reward for the augmented state.
To avoid explicit modeling of reward functions, the key assumption held when performing state augmentations is that the immediate reward for the augmented state $\tilde{\mathbf{s}}_t$ should keep similar to the reward for the original state. 
Therefore, the choice of state augmentation needs to be a local transformation to perturb the original state.
Through performing state augmentations, we are able to artificially let the agent visit states $\tilde{\mathbf{s}}_t$ that do not occur in the offline training data (i.e., $\tilde{\mathbf{s}}_t \in \mathcal{S} \text{ and } \tilde{\mathbf{s}}_t \notin \mathcal{D})$, thus increasing the robustness of the trained RL-agent.

Precisely,  we introduce the following augmentation strategies:
% \begin{figure}[t]
%     \centering
%     \includegraphics[width=1.1\linewidth]{figures/state_augmentation.pdf}
%     \caption{A diagram illustrating the Bellman backup for augmented states. Small perturbations in the original state should not lead to drastically different value functions.}
%     \label{fig:offline RL versus offline RL with state augmentations.}
%     \vspace{-0.6cm}
% \end{figure}

\begin{itemize}[leftmargin=*,nosep]
    \item \header{Zero-mean Gaussian noise}
    Specifically, we add a zero-mean Gaussian noise to the original state as 
    \begin{equation}
        \tilde{\mathbf{s}}_t=\mathbf{s}_t+ \mathbf{\epsilon}_n, \text{where } \mathbf{\epsilon}_n \in \mathcal{N}\left(0, \sigma^2\mathbf{I}\right).
    \end{equation}
    
    $\sigma$ is a hyperparameter to control the variance of Gaussian noise. $\mathbf{I}$ denotes an identity matrix.
    \item \header{Uniform noise}
Similar to the Gaussian noise, we add a uniform noise to the original state as
\begin{equation}
        \tilde{\mathbf{s}}_t=\mathbf{s}_t+ \mathbf{\epsilon}_u, \text{where } \mathbf{\epsilon}_u \in \mathbf{U}\left(\alpha, \beta\right).
    \end{equation}
 $\alpha$ and $\beta$ are hyperparameters to control the noise variance. $\alpha=0.001$
in our experimental settings. 
    \item \header{Item mask}
``Word dropout'' is a common data augmentation technique to avoid overfitting in various natural language processing tasks, such as sentence generation ~\citep{bowman2015generating}, sentiment analysis ~\citep{dai2015semi}, and question answering ~\citep{shubhashri2018lawbo}. It usually masks a random input word with zeros. Inspired by ``Word dropout'', we propose to apply a random item mask as the third augmentation method. 
For each user historical sequence $\mathit{x}_{1:t}$ with $t>T$, we randomly mask one item of the sequence\footnote{We don't apply item mask to sequences whose length is shorter than $T$.}. If the item in the sequence is masked, it is replaced by a special token [mask].

\item \header{Dimension dropout}
Similar to the item mask, we replace one random dimension with zero in the state as $\tilde{\mathbf{s}}_{t}= \mathbf{s}_t \cdot \mathbf{1}$, where $\mathbf{1}$ is a vector of 1s with one 0 randomly sampled from $Bernoulli(p)$. $p$ is a hyperparameter. This transformation preserves all-but-one intrinsic dimension of the original state.
\end{itemize}
We denote a data augmentation transformation as $\mathcal{T}(\tilde{\mathbf{s}}_{\mathit{t}}|\mathbf{s}_{\mathit{t}})$ with $\mathbf{s}_{\mathit{t}} \in \mathcal{D}$. We perform multiple (i.e., $n$ times) augmentations for a given state and 
then the $i$-th step of policy evaluation over the augmented states can be formulated as
\begin{equation}
\begin{aligned}
% \small
    \mathcal{Q}_{\mathit{i+1}}\leftarrow \mathop\text{arg min}_{\substack{\mathcal{Q}}}\mathbb{E}_{\mathcal{D}}
    \sum_{j=1}^n
    [\mathit{r}_{\mathit{t}}&+\gamma\mathcal{Q}_i(\mathcal{T}_{\mathit{j}}(\tilde{\mathbf{s}}_{\mathit{t+1}}|\mathbf{s}_{\mathit{t+1}}),\mathit{a}_{\mathit{t+1}})\\
    &-\mathcal{Q}_i(\mathcal{T}_{\mathit{j}}(\tilde{\mathbf{s}}_{\mathit{t}}|\mathbf{s}_{\mathit{t}}),\mathit{a}_{\mathit{t}})].
\end{aligned}
\end{equation}
And the corresponding augmented loss can be formulated as
\begin{equation}
\label{eq:augmented loss}
    \mathcal{L}_a=-\sum_{j=1}^{n}({\mathit{r}(\mathbf{s}_{\mathit{t}},\mathit{a}_{\mathit{t}})+\gamma\max_{\mathit{a}{\mathit{'}}}\mathcal{Q}(\mathcal{T}_{j}(\tilde{\mathbf{s}}_{\mathit{t+1}}|\mathbf{s}_{\mathit{t+1}}),\mathit{a}{\mathit{'}})-\mathcal{Q}(\mathcal{T}_{\mathit{j}}(\tilde{\mathbf{s}}_{\mathit{t}}|\mathbf{s}_{\mathit{t}}),\mathit{a}_{\mathit{t}})})^2.
\end{equation}
% as shown in Figure \ref{fig:offline RL versus offline RL with state augmentations.}.

The main difference between our proposed objective and the  standard Q-learning is that we augment the Bellman error in Eq. (\ref{eq:q-loss}) to  the sum error over $n$ different augmentations of the same state.
% \textcolor{red}{[A diagram illustrating the difference between a Bellman backup for offline RL versus offline RL with state augmentations is shown in Figure.~\ref{fig:offline RL versus offline RL with state augmentations.}]}
Intuitively, this will help improve the consistency of the Q-value within some perturbation field of the current state $\mathbf{s}_\mathit{t}$, since the Bellman backups are taken over $n$ different views of the same state.
By combining such local perturbations with the states we can distill the benefits of the augmentations to
learn a more robust policy that can generalize better to unseen states when deployed in an online recommendation scenario.  
\subsection{Contrastive Training for RL Recommenders}
\label{subsec:Self-supervision in Offline RL}

We address the issue of large state space and lack of negative signals in implicit feedback. Inspired by the SimCLR framework~\cite{chen2020simple} for learning visual representation, we further propose a reinforcement learning-based contrastive loss to obtain more effective state representations. 
In the supervised settings, the contrastive loss can be minimized by defining the same instance as positive pairs while remaining others as negative pairs. 
While in the RL-based contrastive loss, we define 
%Similarly, the definition of positive pairs in reinforcement learning-based contrastive loss is inspired by instance learning \cite{chen2020improved,he2020momentum,chen2020simple}, 
{different augmentations of the same state as positive pairs. As for negatives, we simply sample another state from another sequence randomly. Hence, the RL-based contrastive loss is to pull the representations of different augmentations of the same sequence state towards each other while pushing the state representations from different sequences apart. }
% The key motivation is to pull the representations of different augmentations of the same sequence state towards each other while pushing the state representations from different sequences apart. 
To achieve this, the contrastive loss learns to minimize the Q-value difference between augmented views of the same state and maximize the Q-value difference between the states derived from other sequences. Specifically, for each state $\mathbf{s}_{t}$ in a mini-batch $\mathit{B}$, we firstly sample another state $\mathbf{s}^{'}_{t}$ from another sequence randomly. Then we apply $\mathit{n}$ same augmentation strategy $\mathcal{T}$ to obtain $\mathit{n}$ different views of $\mathbf{s}_{t}$ (i.e.$\sum_{j=1}^{n}\mathcal{T}_{j}(\tilde{\mathbf{s}}_{\mathit{t}}|\mathbf{s}_{\mathit{t}})$) and average them. Thus the contrastive loss has the form:
\begin{equation}
\label{eq:contrastive loss}
\begin{aligned}
    \mathcal{L}_{\mathit{c}}=-\log\delta
    &((\mathcal{Q}(\mathbf{s}'_{\mathit{t}},\mathit{a}_{\mathit{t}})-\frac{1}{n}\sum_{j=1}^{n}\mathcal{Q}(\mathcal{T}_{\mathit{j}}(\tilde{\mathbf{s}}_{\mathit{t}}|\mathbf{s}_{\mathit{t}}),\mathit{a}_{\mathit{t}}))^2 \\
    &-(\mathcal{Q}(\mathbf{s}_{\mathit{t}},\mathit{a}_{\mathit{t}})-\frac{1}{n}\sum_{j=1}^{n}(\mathcal{Q}(\mathcal{T}_{\mathit{j}}(\tilde{\mathbf{s}}_{\mathit{t}}|\mathbf{s}_{\mathit{t}}),\mathit{a}_{\mathit{t}}))^2), 
\end{aligned}    
\end{equation}
% where $\mathbf{s}'_{\mathit{t}}$  is the sampled negative state for $\mathit{s}_{\mathit{t}}$ and comes from another sequence. 
where $\delta$ is the sigmoid function. Similar contrastive operations can also be conducted for actions.

Finally, we jointly train the weighted sum of the supervised loss, the Q-learning loss, the augmented loss, and the contrastive loss on the offline implicit feedback data:
\begin{equation}
\label{eq:final loss}
    \mathcal{L} = \mathit{w}_{s}\mathcal{L}_{\mathit{s}} + \mathit{w}_{q}\mathcal{L}_{\mathit{q}} +\mathit{w}_{a}\mathcal{L}_{\mathit{a}} + \mathit{w}_{c}\mathcal{L}_{\mathit{c}},
\end{equation}
where $\mathit{w}_{s},\mathit{w}_{q},\mathit{w}_{a},\mathit{w}_{c}$ are weights of the corresponding loss respectively. 
{Without the lose of generality and to avoid  trivial hyperparameter tuning, in our experiments we set all the loss weights to 1 as our default settings.}

\subsection{Discussion}
\label{subsec:discussion}
The proposed CSA can be used as a learning framework and integrated with existing recommendation models, as long as the models can map the input sequence into a hidden state. 
{This keeps inline with most deep learning-based recommendation models introduced over recent years.} 
To verify the performance of CSA, we choose a SOTA recommendation model, i.e., SQN~\cite{xin2020self}, which combines supervised loss and deep Q-learning.
While the proposed CSA can also work for pure RL-based recommendations models. e.g., DQN~\cite{sutton1998introduction} and CQL~\cite{kumar2020conservative}. 
{
 CSA can be seen as an attempt to explore state augmentations and contrastive learning to improve the RL agent trained on the biased state-action data in the offline setting. 
 The proposed methods are for the generic recommendation, i.e., more flexible reward settings such as novelty and dwell time can also be used as the reward function.}

% !TEX root = ../main.tex

\section{Experimental Setup}
\label{sec:Experiments}
In this section, we detail experimental setups to verify the effectiveness of the proposed CSA.
We aim to answer the following research questions:
\begin{enumerate*}[label=\textbf{RQ}\arabic*,leftmargin=*]
\item How does CSA perform in the offline evaluation setting when integrated with existing sequential models? 
\item How does CSA perform when generating multi-round recommendations in the simulated online environment?
\item How does the state augmentation affect CSA performance?
\item {How does contrastive learning affect performance?}
% \item {How does CSA perform when we conduct contrastive operations on state and action respectively?}
\end{enumerate*}

\subsection{Datasets}
We evaluate our method on two publicly accessible datasets: RC15, RetailRocket; and one dataset collected from a real-life serving e-commerce platform:meituan. The RC15 dataset is based on the RecSys Challenge 2015. This dataset is session-based, and each session contains a sequence of clicks and purchases. Following~\citep{xin2020self}, we remove the sessions whose length is smaller than three and then sort the user-item interactions in one session according to the timestamp. 
RetailRocket is collected from a real-world e-commerce website. It contains sequential events of viewing and adding to the cart. {For consistency, we treat views as clicks and adding to the cart as purchases.}

The meituan dataset consists of one week (from April 1st, 2022 to April 7th, 2022) of transaction records on the e-commerce platform. We choose users with both purchases and clicks and then sort the user-item interactions in one session according to the timestamp. We remove items that interacted less than five times and the sequences whose lengths are smaller than 3. 
Table ~\ref{tab:Statistics of datasets.} summarizes the detailed statistics of datasets.
\begin{table}
%   \small
  \centering
  \caption{Statistics of datasets.}
  \label{tab:Statistics of datasets.}
  \begin{tabular}{c ccc}
  \toprule
    Dataset & RC15 & RetailRocket & meituan  \\
  \midrule
    \#sequences & 200,000 & 195,523 & 108,650 \\
    \#items  & 26,702 & 70,852 & 87,987  \\
    \#clicks & 1,110,965 & 1,176,680 & 1,966,922  \\
    \#purchases & 43,946 & 57,269 & 427,701  \\
  \bottomrule
\end{tabular}
\vspace{-0.4cm}
\end{table}

\subsection{Evaluation Metrics}
We employ top-$\mathit{k}$ Hit Ratio (HR@$\mathit{k}$), top-$\mathit{k}$ Normalized Discount Cumulative Gain (NDCG@$\mathit{k}$) to evaluate the performance, which is widely used in related works~\citep{rendle2010factorizing,xin2020self}. HR@$\mathit{k}$ is a recall-based metric, measuring whether the ground-truth item is in the top-$k$ positions of the recommendation list. NDCG is a rank-sensitive metric which assigns higher scores to top positions in the recommendation list. We report results of  HR@\{5,10,20\}, NDCG@\{5,10,20\}. For the two publicly accessible datasets, we use the sample data splits with ~\citep{xin2020self}. For the anonymous dataset, the training, validation, and test set ratio is 8:1:1.
The ranking is performed among the whole item set. Note that we use both clicks and purchases for model training and report the HR and NDCG on purchase predictions since this work focuses on generating recommendations that can lead to high-reward user feedback.

\subsection{Baselines}
\label{subsec:baselines}
We integrate the proposed CSA with three state-of-the-art sequential recommendation models:
\begin{enumerate*}
\item {\textbf{GRU4Rec}~\citep{hidasi2015session}} is a classical RNN-based method for a session-based recommendation.%; it stacks multiple GRU layers for session-parallel mini-batch training.
\item{\textbf{SASRec}~\citep{kang2018self}} is a self-attention-based sequential recommendation model, which uses the multi-head attention mechanism to recommend the next item.
\item{\textbf{FMLPRec}~\citep{zhou2022filter}} stacks multiple multi-layer perceptrons (MLP) blocks to produce the representation of sequential user preference for the recommendation.
\end{enumerate*}

Each model is trained using the following methods:
\begin{enumerate*}
\item \textbf{Normal} denotes training the model with the normal supervised cross-entropy loss.%; it stacks multiple GRU layers for session-parallel mini-batch training.
\item \textbf{SQN} \cite{xin2020self} combines the supervised loss and deep Q-learning.
\item \textbf{CSA-N} denotes CSA with the Gaussian noise augmentation.
\item \textbf{CSA-U} is  CSA with the uniform noise augmentation.
\item \textbf{CSA-M} is CSA with the item mask augmentation.
\item \textbf{CSA-D} denotes CSA with the dimension dropout augmentation.
\end{enumerate*}

\subsection{Implementation details}
\label{subsec:implementation details}
For all datasets, the input sequences are composed of the last ten items before the target timestamp. 
We complement the sequence with a padding item if the sequence length is less than 10. We use a mini-batch Adam optimizer to optimize the model. The batch size is set as 256. For a fair comparison, the embedding size is set as 64 for all models. The learning rate is tuned to the best for a given model and dataset.
Specifically, for GRU4Rec and SASRec, the learning rate is set as 0.01 for RC15, 0.005 for RetailRocket and 0.001 for the anonymous dataset. The number of heads in self-attention of SASRec is set as 1 according to its original paper~\citep{kang2018self}. For FMLPRec, the learning rate is set as 0.001 for all datasets. %The number of heads and blocks are set as 1, 1 respectively. 
For CSA, the times of state augmentations are set as $n=2$\footnote{Our code used in this work is available at \changed{\url{https://github.com/HN-RS/CSA}}.}.

% !TEX root = ../main.tex

\section{Experimental Results}
\label{sec:Results}
In this section, we present the experimental results to answer the research questions described in section \ref{sec:Experiments}.
\begin{table*}
	\centering
	\caption{Offline performance comparison of different methods on RC15 and RetailRocket. Boldface denotes the highest score. HR and NG are short for Hit Ratio and NDCG,  respectively.}
    \begin{adjustbox}{max width=\linewidth}
	\begin{tabular}{cc cccccc cccccc}
	\toprule
	\multicolumn{2}{c}{\multirow{2}*{method}} &\multicolumn{6}{c}{RC15} &\multicolumn{6}{c}{RetailRocket}\\
	\cmidrule(lr){3-8} \cmidrule(lr){9-14} 
	 & & HR@5 & NG@5 & HR@10 & NG@10  & HR@20 & NG@20 & HR@5 & NG@5 & HR@10 & NG@10  & HR@20 & NG@20 \\
	\toprule
	\multirow{6}{*}{GRU4Rec}
    &normal             &0.4057 &0.2802 &0.5191 &0.3170 &0.6093 &0.3398  
                        &0.4559 &0.3775 &0.5092 &0.3948 &0.5551 &0.4064 
                          \\ 
    &SQN         &0.4313 &0.3043 &0.5433 &0.3405 &0.6332 &0.3634  
                        &0.5009 &0.4135 &0.5538 &0.4308 &0.5993 &0.4423 
                          \\ 
    \cmidrule(lr){2-14}
    
    &CSA-N &0.4463 &0.3250 &0.5578 &0.3613 &0.6436 &0.3830
           &0.5111 &0.4298 &0.5608 &0.4459 &\textbf{0.6020} &0.4563 
                          \\ 
   
    &CSA-U &0.4454 &0.3224 &0.5558 &0.3582 &0.6427 &0.3803
           &0.5075 &0.4264 &0.5594 &0.4434 &0.5989 &0.4534
                         \\ 
   
    &CSA-M &\textbf{0.4525} &\textbf{0.3268} &\textbf{0.5629} &\textbf{0.3627} &\textbf{0.6531} &\textbf{0.3856}       
           &\textbf{0.5126} &\textbf{0.4358} &\textbf{0.5638} &\textbf{0.4524} &0.6018 &\textbf{0.4621}
                         \\ 
    &CSA-D &0.4435 &0.3174 &0.5604 &0.3555 &0.6466 &0.3774
           &0.5103 &0.4256 &0.5583 &0.4412 &0.6010 &0.4520
                         \\ 
	\toprule
	\multirow{6}{*}{SASRec}
	&normal             &0.4308 &0.3006 &0.5512 &0.3396 &0.6385 &0.3619 
	                   &0.5335 &0.4322 &0.5878 &0.4497 &0.6330 &0.4611
	                   \\
	&SQN         &0.4433 &0.3098 &0.5509 &0.3448 &0.6427 &0.3681
	                   &0.5649 &0.4748 &0.6203 &0.4929 &0.6545 &0.5016
	                   \\
	\cmidrule(lr){2-14}
	&CSA-N       &0.4537 &0.3208 &\textbf{0.5705} &0.3587 &\textbf{0.6598} &0.3812
	             &\textbf{0.5874} &\textbf{0.4934} &0.6316 &0.5078 &0.6719 &0.5180
	                   \\ 
	&CSA-U       &0.4364 &0.3122 &0.5470 &0.3482 &0.6346 &0.3705
	             &0.5855 &0.4914 &\textbf{0.6390} &\textbf{0.5090} &\textbf{0.6768} &\textbf{0.5185}
	                   \\ 
    &CSA-M    &\textbf{0.4550} &\textbf{0.3243} &0.5701 &\textbf{0.3617} &0.6535 &\textbf{0.3828} 
                       &0.5683 &0.4745 &0.6269 &0.4937 &0.6693 &0.5044
                       \\ 
    &CSA-D    &0.4509 &0.3184 &0.5669 &0.3561 &0.6574 &0.3792
                       &0.5738 &0.4773 &0.6280 &0.4950 &0.6630 &0.5038
                       \\ 
    
	\toprule
	\multirow{6}{*}{FMLPRec}
	&normal            &0.4087 &0.2929 &0.5145 &0.3272 &0.6010 &0.3493
	                   &0.5218 &0.4307 &0.5712 &0.4468 &0.6118 &0.4571
	                   \\
	&SQN        &0.4205 &0.2994 &0.5304 &0.3353 &0.6199 &0.3581 
	                   &0.5534 &0.4631 &0.5991 &0.4781 &0.6366 &0.4876
	                   \\
	\cmidrule(lr){2-14}
	&CSA-N      &\textbf{0.4394} &\textbf{0.3209} &0.5422 &\textbf{0.3543} &0.6251 &\textbf{0.3752} 
	                   &\textbf{0.5876} &0.5041 &0.6313 &0.5183 &\textbf{0.6696} &0.5280
	                   \\
	&CSA-U      &0.4359 &0.3125 &0.5420 &0.3472 &\textbf{0.6277} &0.3690 
	                   &0.5870 &\textbf{0.5078} &\textbf{0.6320} &\textbf{0.5223} &0.6675 &\textbf{0.5312}
	                   \\
	&CSA-M   &0.4359 &0.3140 &\textbf{0.5436} &0.3489 &0.6277 &0.3703 
	                   &0.5868 &0.5004 &0.6297 &0.5144 &0.6679 &0.5242
	                   \\
	&CSA-D &0.4391 &0.3173 &0.5422 &0.3508 &0.6268 &0.3722
	                    &0.5816 &0.4940 &0.6254 &0.5083 &0.6611 &0.5147
	                    \\
	\bottomrule
	\end{tabular}
	\end{adjustbox}
	\label{tab:Overall Performance on RC15 and RetailRocket.}
	
\end{table*}
\begin{table}
\vspace{-0.1cm}
	\centering
	% \caption{Offline performance comparison of methods on meituan dataset.}
    \caption{Offline performance comparison of methods on meituan dataset. Boldface denotes the highest score. HR and NG are short for Hit Ratio and NDCG, respectively.}
    \begin{adjustbox}{max width=\linewidth}
	\begin{tabular}{cc cccccc}
	\toprule
	\multicolumn{2}{c}{\multirow{2}*{method}} &\multicolumn{6}{c}{meituan}\\
	\cmidrule(lr){3-8}
	 & & HR@5 & NG@5 & HR@10 & NG@10  & HR@20 & NG@20 \\
    
	\toprule
	\multirow{6}{*}{GRU4Rec}
    & normal           &0.5139 &0.4352 &0.5676 &0.4526 &0.6110 &0.4636  \\ 
    & SQN                &0.5238 &0.4484 &0.5783 &0.4661 &0.6224 &0.4773  \\ 
    \cmidrule(lr){2-8}
    &CSA-N       
                        &0.5256 &0.4528 &\textbf{0.5799} &0.4706 &\textbf{0.6242} &0.4818  \\ 
    &CSA-U     
                        &\textbf{0.5270} &\textbf{0.4565} &0.5795 &\textbf{0.4735} &0.6220 &\textbf{0.4843} \\ 
    &CSA-M    
                        &0.5095 &0.4385 &0.5618 &0.4554 &0.6066 &0.4668 \\ 
    &CSA-D 
                        &0.5245 &0.4502 &0.5774 &0.4674 &0.6203 &0.4783 \\ 
	\toprule
	\multirow{6}{*}{SASRec}
	&normal             
	                   &0.5230 &0.4299 &0.5816 &0.4489 &0.6272 &0.4605\\
	&SQN        
	                   &0.5352 &0.4416 &0.5952 &0.4612 &0.6398 &0.4725\\ 
	\cmidrule(lr){2-8}               
	&CSA-N       
	                   &\textbf{0.5455} &\textbf{0.4595} &0.6016 &\textbf{0.4778} &\textbf{0.6482} &\textbf{0.4896}\\ 
	&CSA-U       
	                   &0.5453 &0.4560 &\textbf{0.6040} &0.4751 &0.6472 &0.4861\\ 
    &CSA-M    
                       &0.5331 &0.4412 &0.5908 &0.4600 &0.6368 &0.4717\\ 
    &CSA-D 
                       &0.5366 &0.4432 &0.5946 &0.4620 &0.6408 &0.4737\\ 
    
	\toprule
	\multirow{6}{*}{FMLPRec}
	&normal            
	                   &0.5132 &0.4310 &0.5721 &0.4501 &0.6166 &0.4614\\
	&SQN        
	                   &0.5316 &0.4492 &0.5877 &0.4675 &0.6313 &0.4786\\
	\cmidrule(lr){2-8}                   
	&CSA-N      
	                   &0.5408 &\textbf{0.4663} &\textbf{0.5935} &\textbf{0.4833} &0.6359 &0.4941\\
	&CSA-U      
	                   &\textbf{0.5415} &0.4659 &0.5926 &0.4825 &\textbf{0.6395} &\textbf{0.4944}\\
	&CSA-M   
	                   &0.5308 &0.4517 &0.5862 &0.4697 &0.6306 &0.4809\\
	&CSA-D 
	                    &0.5339 &0.4627 &0.5888 &0.4806 &0.6313 &0.4914\\
	\bottomrule
	\end{tabular}
	\end{adjustbox}
	\label{tab:Overall Performance on Anonymous.}
	\vspace{-0.3cm}
\end{table}

\subsection{Offline Performance Comparison (RQ1)}
\label{subsec:Overall Performance Comparison}
Table~\ref{tab:Overall Performance on RC15 and RetailRocket.} and Table~\ref{tab:Overall Performance on Anonymous.} show the offline performance comparison on the two {publicly} accessible datasets and the anonymous dataset, respectively. 

We have the following observations:
\begin{enumerate*}[label=(\roman*)]
\item Compared with the standard supervised training, it is evident that RL-based training methods perform better on all datasets, demonstrating that incorporating RL can help improve the recommendation performance.
\item Our proposed CSA further improves the performance over SQN with almost all state augmentation strategies. On RC15 and RetailRocket, four data augmentation strategies help consistently to improve the performance, except for CSA-U with the SASRec model on RC15. On the anonymous dataset, CSA-N, CSA-U and CSA-D consistently outperform the SQN method, while CSA-M performs comparably with SQN. For cases in which CSA cannot outperform the SQN method, the reason could be that the augmented noises are too aggressive so the value functions could be changed for the augmented states.
\end{enumerate*}
To conclude, the proposed CSA is effective to improve the RL-based recommendation performance in the offline evaluation setting. 

\subsection{Online Performance Comparison (RQ2)}
\label{subsec:Online Performance}
\subsubsection{Kuaishou Environment}
{KuaiRec is a real-world dataset collected from the recommendation logs of a video-sharing mobile app. 
}
There are 1,411 users and 3,327 videos that compose a matrix where each entry represents a user’s feedback on a video. 
The density of this matrix is almost 100\% (i.e., 99.6\%), which means that we can always
get a user's feedback for every item. 
Such a dataset can be regarded as a simulated online environment since we can get a reward for every possible action.

\subsubsection{Implementation and Evaluation}
{In this environment, we conduct 10 rounds of recommendation to each user and average the obtained rewards of all users to get the immediate reward in each round. The reward is defined as the watching ratio of a video. }
Since the objective of RL is to maximize the cumulative gains in the long-run, we report the discounted accumulative reward of the 10-round recommendation process. Experiments are conducted three times, and the average is reported.

\begin{figure}
\vspace{-0.1cm}
    \centering
    \includegraphics[width=0.75\linewidth]{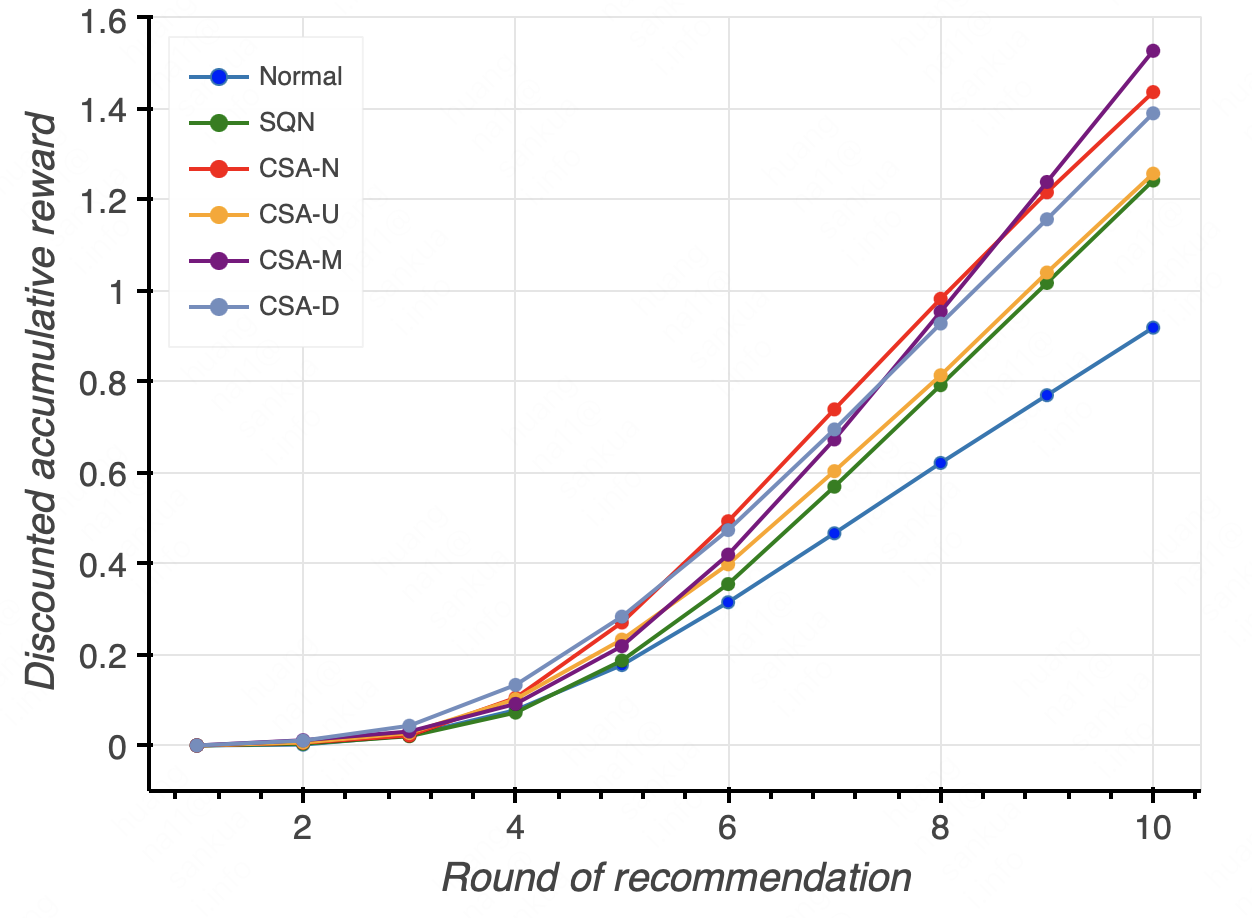}
    \caption{Comparison of multi-round recommendation performance. GRU4Rec is the base sequential model.}
    \label{fig: online performance.}
    \vspace{-0.5cm}
\end{figure}

\subsubsection{Performance Analysis}
The performance comparison of multi-round recommendation is illustrated in Figure~\ref{fig: online performance.}. We choose GRU4Rec as the base sequential model.
It is evident that our proposed CSA significantly outperforms
standard training and SQN.
% other baseline models. 
Besides, we can observe more insightful facts:
\begin{enumerate*}[label=(\roman*)]
\item in comparison of the accumulated reward between normal supervised training and RL-based training, we find that with the number of recommendation rounds increasing, the RL-based methods can achieve more significant gains than the supervised method, indicating the intrinsic advantage of RL to {maximize} long-term benefits;
\item compared with SQN, the proposed methods achieve higher accumulative rewards in each round of recommendation, demonstrating the effectiveness of the state augmentations;
\item when comparing the performance of four state augmentation methods, we find that CSA-N achieves the highest reward in the first nine {rounds. In contrast, CSA-M achieves the highest reward in the tenth round, implying that CSA-M may perform better for long sequences. }
\end{enumerate*}

\subsection{Effect of State Augmentations (RQ3)}
\label{subsec:Influence of state augmentation}
In this section, we first examine how the number of state augmentations (i.e., $n$) affects the method performance.
Then we investigate how the hyperparameters of state augmentation methods influence their performance. 
%Due to the space limitation, 
We report the experimental results on the RC15 dataset with GRU4Rec as the sequential model. Results on other datasets and sequential models show similar trends.
\subsubsection{Effect of the number of state augmentations}
\label{subsec:the number of state augmentation}
\begin{figure}[t]
\vspace{-0.1cm}
    \centering
    \includegraphics[width=1.0\linewidth]{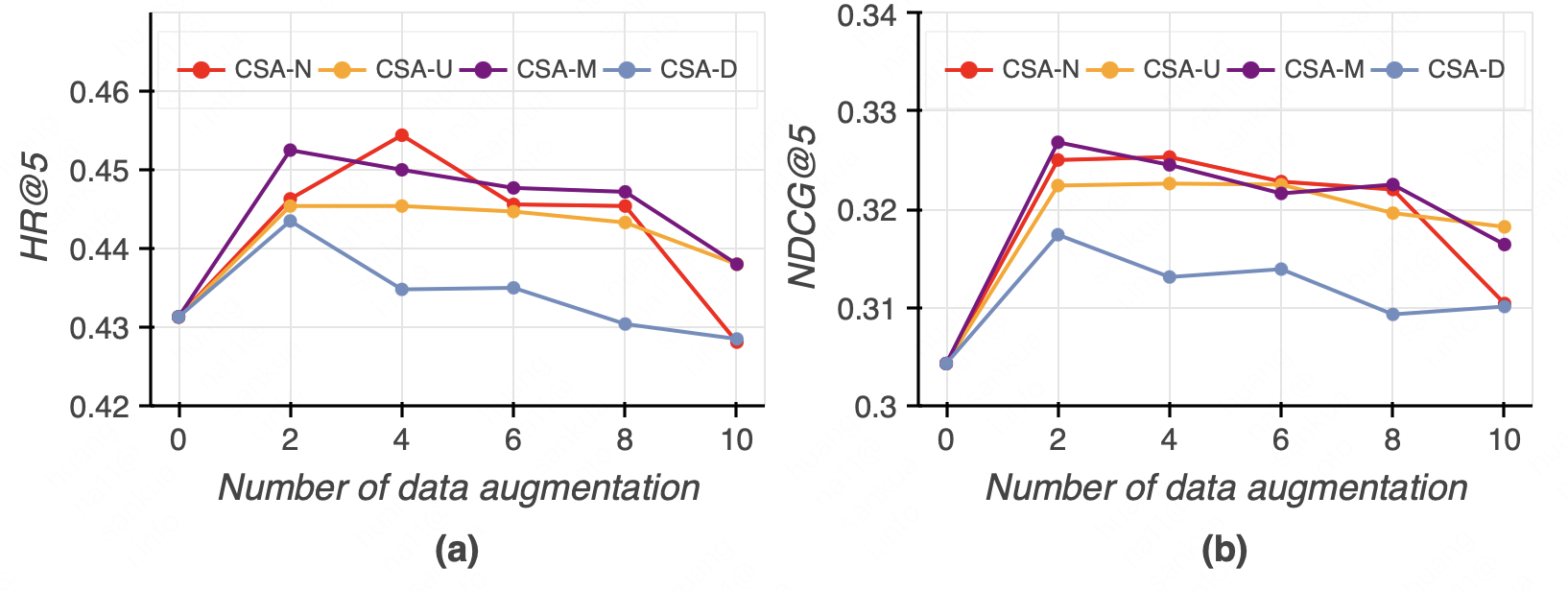}
    \caption{Effect of the number of state augmentations.}
    \label{fig: hyper_na.}
    \vspace{-0.5cm}
\end{figure}

The results of HR@5 and NDCG@5 are shown in Figure~\ref{fig: hyper_na.}(a) and Figure~\ref{fig: hyper_na.}(b) respectively. We can see that when the number of state augmentations is greater than 0, the performance is significantly improved, which demonstrates the effectiveness of the state augmentation strategies. 

Considering the performance and efficiency, the number of state augmentation is set to 2 in our experiments. 
{In addition, we can observe that method performance starts to decrease with a further increase in the augmentation number. The reason could be that too many augmentations would make more states have similar Q-values, thus increasing the risk of model collapse.}

\subsubsection{Effect of state augmentation hyperparameters}
\label{subsubsec:various hyper parameters}
\begin{figure}[t]
    \centering
    \includegraphics[width=1.0\linewidth]{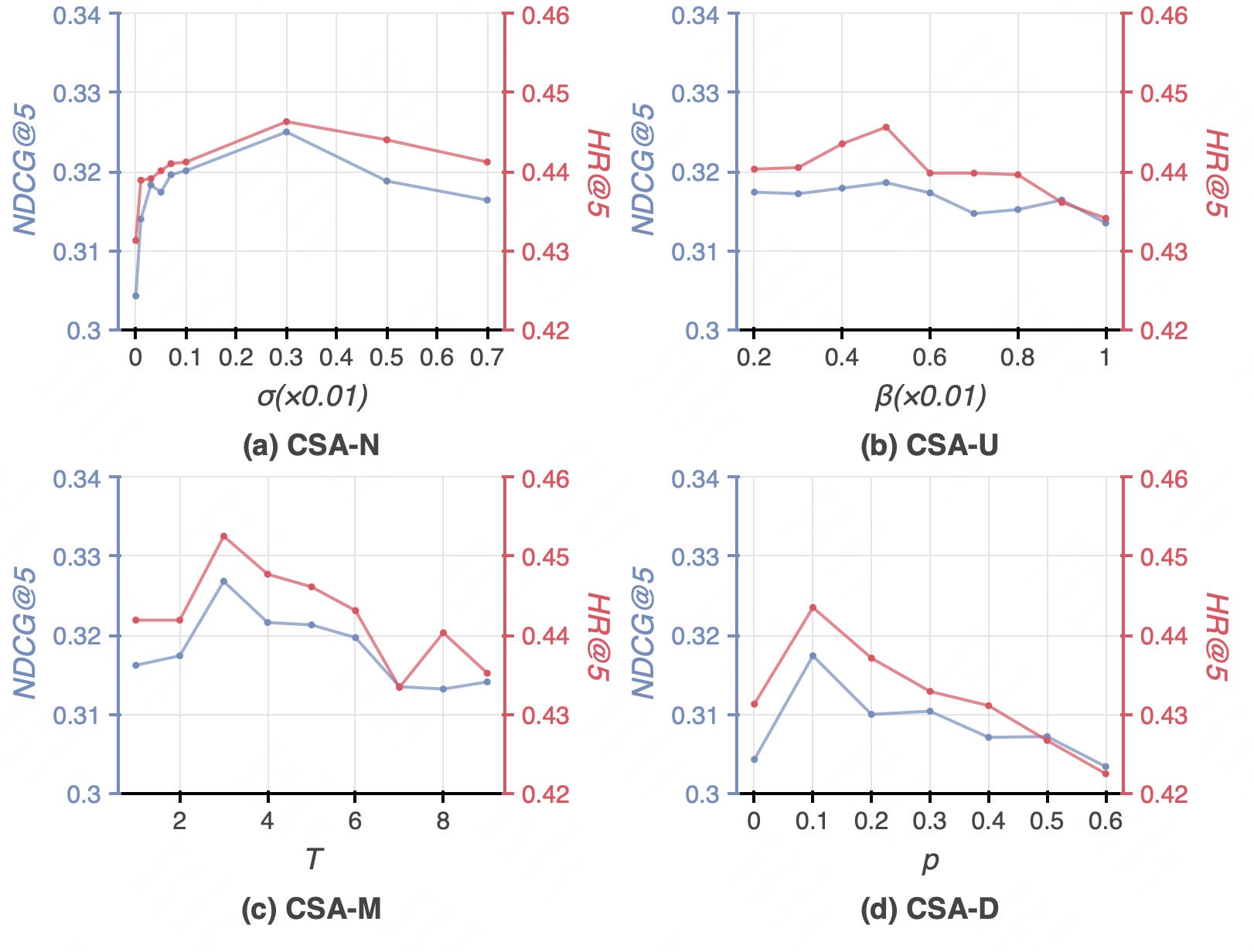}
    \caption{Effect of state augmentation hyperparameters.}
    \label{fig: hyper.}
\end{figure}

We investigate the performance sensitivity regarding to $\sigma$ in CSA-N, $\beta$ in CSA-U, $T$ in CSA-M, and $\mathit{p}$ in CSA-D. The results are illustrated in Figure~\ref{fig: hyper.}(a), Figure~\ref{fig: hyper.}(b), Figure~\ref{fig: hyper.}(c) and Figure~\ref{fig: hyper.}(d) respectively. We can see that the performance of CSA-N and CSA-U first improves with the increase of the $\sigma$ and $\beta$, which are hyperparameters used to control the magnitude of the noise.
{The best performance is achieved when $\sigma=0.003$ and $\beta=0.005$, indicating that the noises introduced at such values effectively perturb the state while maintaining the original semantic information. Then as the introduced noises become larger, the performance starts to decrease, demonstrating that too aggressive noises would also hurt the method's performance. }

{For CSA-M, the best performance is achieved when $T=3$. On the one hand, masking an item could lead to a dramatic state change with poor model performance when $T<3$. }On the other hand,  $T>3$ also decreases the performance since a larger $T$ indicates a smaller portion of states will be augmented.
 
For CSA-D, we can find that the best performance is achieved when $p=0.1$.

\subsection{Effect of contrastive learning (RQ4)}
\label{subsec:Effect of contrastive learning(RQ4)}

\subsubsection{Ablation study of the contrastive loss.}
{This section examines how the contrastive learning term (i.e., $\mathcal{L}_c$) contributes to the method performance. 
The results of GRU4Rec and SASRec on the RC15 dataset are shown in Table~\ref{tab:Ablation studies.}. We can see that when the contrastive loss is removed, the method performance drops in almost all cases. 
Such results demonstrate the proposed contrastive term's effectiveness in improving the recommendation performance. 
The contrastive loss helps to learn more effective state representations by making the Q-values from different augmentations over one sequence state similar while pushing away the Q-values over the state coming from another sequence.}

\begin{table}
	\centering
	\caption{Effect of $\mathcal{L}_c$. Boldface denotes the higher score. HR and NG are short for Hit Ratio and NDCG, respectively. GRU and SAS are short for GRU4Rec and SASRec. N,U,M,D denote four augmentation strategies. -c denotes the removal of $\mathcal{L}_c$.}
    \begin{adjustbox}{max width=\linewidth}
	\begin{tabular}{cc cccccc}
	\toprule
	\multicolumn{2}{c}{\multirow{2}*{method}} &\multicolumn{6}{c}{RC15}\\
	\cmidrule(lr){3-8}
	 & & HR@5 & NG@5 & HR@10 & NG@10  & HR@20 & NG@20 \\
    
	\toprule
	\multirow{8}{*}{GRU}
    & N     &\textbf{0.4463} &\textbf{0.3250} &\textbf{0.5578} &\textbf{0.3613} &\textbf{0.6436} &\textbf{0.3830}  \\ 
    & N-c   &0.4463 &0.3201 &0.5526 &0.3546 &0.6335 &0.3751  \\ 
    \cmidrule(lr){2-8}
    &U      &0.4454 &\textbf{0.3224} &\textbf{0.5558} &\textbf{0.3582} &\textbf{0.6427} &\textbf{0.3803}  \\ 
    &U-c    &\textbf{0.4458} &0.3173 &0.5535 &0.3521 &0.6427 &0.3748 \\ 
    \cmidrule(lr){2-8}
    &M    &\textbf{0.4525} &\textbf{0.3268} &\textbf{0.5629} &\textbf{0.3627} &\textbf{0.6531} &\textbf{0.3856} \\ 
    &M-c  &0.4477 &0.3216 &0.5546 &0.3563 &0.6429 &0.3787 \\ 
    \cmidrule(lr){2-8}
    &D    &\textbf{0.4435} &\textbf{0.3174} &\textbf{0.5604} &\textbf{0.3555} &\textbf{0.6466} &\textbf{0.3774} \\ 
    &D-c  &0.4431 &0.3141 &0.5516 &0.3494 &0.6383 &0.3713\\ 
	\toprule
	\multirow{8}{*}{SAS}
    & N     &\textbf{0.4537} &\textbf{0.3208} &0.5705 &0.3587 &\textbf{0.6598} &\textbf{0.3812}  \\ 
    & N-c   &0.4532 &0.3207 &\textbf{0.5735} &\textbf{0.3599} &0.6577 &0.3812  \\ 
    \cmidrule(lr){2-8}
    &U      &\textbf{0.4364} &\textbf{0.3122} &\textbf{0.5470} &\textbf{0.3482} &\textbf{0.6346} &\textbf{0.3705}  \\ 
    &U-c    &0.4085 &0.2887 &0.5053 &0.3200 &0.5876 &0.3409 \\ 
    \cmidrule(lr){2-8}
    &M      &\textbf{0.4550} &\textbf{0.3243} &\textbf{0.5701} &\textbf{0.3617} &\textbf{0.6535} &\textbf{0.3828}  \\ 
    &M-c    &0.4424 &0.3149 &0.5581 &0.3523 &0.6455 &0.3745  \\ 
    \cmidrule(lr){2-8}
    &D      &\textbf{0.4509} &\textbf{0.3184} &\textbf{0.5669} &\textbf{0.3561} &\textbf{0.6574} &\textbf{0.3792}  \\  
    &D-c    &0.4449 &0.3119 &0.5535 &0.3473 &0.6475 &0.3713  \\ 
    
	\bottomrule
	\end{tabular}
	\end{adjustbox}
	\label{tab:Ablation studies.}
	\vspace{-0.5cm}
\end{table}

\subsubsection{Comparison of contrastive states and contrastive actions.}

In this section, we explore the comparison between contrastive states and contrastive actions. The results of GRU4Rec and SASRec on the RC15 dataset are shown in Table~\ref{tab:Ablation studies1.}. 
Combining the results of Table ~\ref{tab:Ablation studies1.} and Table \ref{tab:Overall Performance on RC15 and RetailRocket.}, we can see that both the contrastive states and contrastive actions can help to improve the recommendation performance, obtaining better results than SQN. This observation demonstrates that introducing contrastive signals can help to improve RL-based recommender systems. 
Besides, we also find that in most cases, {contrastive operations conducted on actions perform worse than that conducted on states. The reason is that different models have different encoding capabilities to map the input sequences into hidden states. As a result, the contrastive operations conducted on states tend to help the recommender to learn more stable state representations, leading to better results.}

\begin{table}
	\centering
	% \caption{Effect of contrastive operations conducted on states(s) and actions(a).}
    \caption{Effect of contrastive operations conducted on states and actions. Boldfae denotes the higher score. HR and NG are short for Hit Ratio and NDCG, respectively. GRU and SAS are short for GRU4Rec and SASRec. N,U,M,D denote four augmentation strategies. (s) denotes contrastive operations are conducted on states. (a) denotes contrastive opertations are conducted on actions.}
    \begin{adjustbox}{max width=\linewidth}
	\begin{tabular}{cc cccccc}
	\toprule
	\multicolumn{2}{c}{\multirow{2}*{method}} &\multicolumn{6}{c}{RC15}\\
	\cmidrule(lr){3-8}
	 & & HR@5 & NG@5 & HR@10 & NG@10  & HR@20 & NG@20 \\
    
	\toprule
	\multirow{8}{*}{GRU}
    & N(s)     &\textbf{0.4463} &\textbf{0.3250} &\textbf{0.5578} &\textbf{0.3613} &\textbf{0.6436} &\textbf{0.3830}  \\ 
    & N(a)   &0.4375 &0.3152 &0.5470 &0.3508 &0.6319 &0.3724  \\ 
    \cmidrule(lr){2-8}
    &U(s)      &\textbf{0.4454} &\textbf{0.3224} &\textbf{0.5558} &\textbf{0.3582} &\textbf{0.6427} &\textbf{0.3803}  \\ 
    &U(a)    &0.4391 &0.3151 &0.5503 &0.3513 &0.6381 &0.3736 \\ 
    \cmidrule(lr){2-8}
    &M(s)    &\textbf{0.4525} &\textbf{0.3268} &\textbf{0.5629} &\textbf{0.3627} &\textbf{0.6531} &\textbf{0.3856} \\ 
    &M(a)  &0.4419 &0.3162 &0.5565 &0.3535 &0.6408 &0.3749 \\ 
    \cmidrule(lr){2-8}
    &D(s)    &\textbf{0.4435} &\textbf{0.3174} &\textbf{0.5604} &\textbf{0.3555} &\textbf{0.6466} &\textbf{0.3774} \\ 
    &D(a)  &0.4320 &0.3053 &0.5392 &0.3399 &0.6289 &0.3626\\ 
	\toprule
	\multirow{8}{*}{SAS}
    & N(s)     &\textbf{0.4537} &0.3208 &\textbf{0.5705} &0.3587 &0.6598 &0.3812  \\ 
    & N(a)   &0.4532 &\textbf{0.3238} &0.5701 &\textbf{0.3620} &\textbf{0.6600} &\textbf{0.3849}  \\ 
    \cmidrule(lr){2-8}
    &U(s)      &\textbf{0.4364} &\textbf{0.3122} &\textbf{0.5470} &\textbf{0.3482} &\textbf{0.6346} &\textbf{0.3705}  \\ 
    &U(a)   &0.4101 &0.2934 &0.5129 &0.3268 &0.5959 &0.3478 \\ 
    \cmidrule(lr){2-8}
    &M(s)      &\textbf{0.4550} &\textbf{0.3243} &\textbf{0.5701} &\textbf{0.3617} &\textbf{0.6535} &\textbf{0.3828}  \\ 
    &M(a)    &0.4507 &0.3190 &0.5611 &0.3549 &0.6500 &0.3775  \\ 
    \cmidrule(lr){2-8}
    &D(s)      &\textbf{0.4509} &\textbf{0.3184} &\textbf{0.5669} &\textbf{0.3561} &\textbf{0.6574} &\textbf{0.3792}  \\  
    &D(a)    &0.4461 &0.3167 &0.5558 &0.3525 &0.6438 &0.3749  \\ 
    
	\bottomrule
	\end{tabular}
	\end{adjustbox}
	\label{tab:Ablation studies1.}
	\vspace{-0.6cm}
\end{table}
% !TEX root = ../main.tex

\section{Conclusion and future work}
\label{sec:Conclusion}
{This paper proposed contrastive state augmentation (CSA) for training RL-based recommender systems to address the issues of inaccurate value estimation for unseen states and insufficient state representation learning from implicit user feedback.}
We devised four state augmentation strategies to improve the generalization capability of the RL recommendation agent. In addition, we introduced contrastive signals to facilitate state representation learning. To verify the effectiveness of CSA, we implemented it with three state-of-the-art sequential recommendation models. We conducted extensive experiments on three real-world datasets and one simulated online environment.
Experimental results demonstrate that the proposed CSA training framework can effectively improve  recommendation performance. In the future, we intend to investigate more state augmentation strategies and design more effective contrastive loss to  improve the performance further. We believe that combining self-supervised representation learning and RL could be a promising direction for future recommender systems.
\vspace{-0.2cm}

\begin{acks}
This research was funded by the Natural Science Foundation of China (62272274,61972234,62072279,62102234,62202271), Meituan, the Natural Science Foundation of Shandong Province (ZR2021QF129), the Key Scientific and Technological Innovation Program of Shandong Province (2019JZZY010129), Shandong University multidisciplinary research and innovation team of young scholars (No.2020QNQT017), the Tencent WeChat Rhino-Bird Focused Research Program (JR-WXG2021411), the Fundamental Research Funds of Shandong University.
\end{acks}

\bibliographystyle{ACM-Reference-Format}
\balance
\bibliography{main}

\clearpage
\end{sloppypar}
\end{document}